\documentclass[twocolumn,secnumarabic,amssymb, nobibnotes, aps, pra]{revtex4-1}

\usepackage{braket}
\usepackage{graphicx}
\usepackage{color,soul}
\usepackage{amsmath}
\usepackage{amssymb}

\begin{document}

\title{Temporal-mode measurement tomography of a quantum pulse gate}

\author{Vahid Ansari$^{1}$}
\email{vahid.ansari@uni-paderborn.de}
\author{Georg Harder$^{1}$}
\author{Markus Allgaier$^{1}$}
\author{Benjamin Brecht$^{1,2}$}
\author{Christine Silberhorn$^{1}$} 
\affiliation{$^1$Integrated Quantum Optics, Paderborn University, Warburger Strasse 100, 33098 Paderborn, Germany}
\affiliation{$^2$Clarendon Laboratory, Department of Physics, University of Oxford, Parks Road, OX1 3PU, United Kingdom}

\begin{abstract}
Encoding quantum information in the photon temporal mode (TM) offers a robust platform for high-dimensional quantum protocols. The main practical challenge, however, is to design a device that operates on single photons in specific TMs and all coherent superpositions. The quantum pulse gate (QPG) is a mode-selective sum-frequency generation designed for this task. Here, we perform a full modal characterisation of a QPG using weak coherent states in well-defined TMs. We reconstruct a full set of measurement operators, which show an average fidelity of 0.85 to a theoretically ideal device when operating on a 7-dimensional space. Then we use these characterised measurement operators of the QPG to calibrate the device. Using the calibrated device and a tomographically complete set of measurements, we show that the QPG can perform high-dimensional TM state tomography with 0.99 fidelity.
\end{abstract}
\date{\today}

\maketitle

\section{Introduction}
Optical quantum information science (QIS) covers a multitude of applications ranging from quantum computing and simulation over quantum metrology to quantum communications. Using photons to carry information in any of these applications, we have to choose an alphabet for information encoding. Of the four degrees of freedom --- polarisation, transverse electric field distribution (two degrees of freedom) and time --- polarisation is more popular due to its experimental accessibility. This comes, however, with an intrinsic limitation to a two-dimensional Hilbert space, where we actually would prefer an infinite-dimensional alphabet which can increase the information capacity of each photon and can also improve the performance of quantum protocols. 
For this reason, recent years have seen increasing interest in alternative encodings deploying either the spatial degree of freedom or the spectral-temporal domain where the basis states are e.g. orbital angular momentum states or temporal modes (TMs), respectively. The latter are particularly appealing because they are compatible with single-mode fibre networks and are also Eigenmodes of state of the art photon sources based on parametric down-conversion and four-wave mixing. However, the temporal shaping and detection of single-photon wave packets in higher-dimensional spaces is challenging, as it requires time-dependant operations, such as nonlinear optical interactions \cite{Brecht2011a,Eckstein2011}. Regardless of this, TMs of single photons have been identified as a promising resource for QIS and were studied in many contexts such as: high-dimensional quantum communications \cite{Brecht2015}, deterministic photonic quantum gates \cite{Ralph2015}, light-matter interaction \cite{Nunn2007,Zheng2015}, and enhanced-resolution spectroscopy \cite{Schlawin2016}. 
Any of these applications necessarily require the capability to prepare photons in specific TMs, defined by a complex amplitude and phase distribution of the electric field, and to perform TM-resolved measurements in both the computational and any associated superposition basis. This can be achieved with the quantum pulse gate (QPG), a device that selects a single, arbitrary TM and converts it to a distinguishable output \cite{Brecht2011a,Eckstein2011}. Recently, such devices have been demonstrated by several groups by employing dispersion-engineered frequency conversion between a strong shaped driving pump field and a coherent signal state at single-photon level intensities \cite{Brecht2014,Manurkar2016,Ra2017,Shahverdi2017,Reddy2017} or with heralded single photons from a parametric down-conversion source \cite{Ansari2016}. In these experiments, TM selectivity with reasonable efficiencies has been shown, but the coherences between all possible TMs have not been investigated in detail. This is, however, an essential ingredient for the realisation of any application based on a high-dimensional alphabet rather than on simple add/drop-type multiplexing of information channels. An easy example is polarisation tomography, where measurements have to be carried out in all three mutually unbiased bases (MUBs) --- horizontal/vertical, diagonal/anti-diagonal, right-circular/left-circular --- in order to retrieve full information on the state under investigation. 

In this work, we reconstruct all measurement operators of a QPG operating on both a 5-dimensional and 7-dimensional TM Hilbert space. Our QPG is based on dispersion-engineered sum-frequency generation in a titanium-indiffused lithium niobate waveguide, and we use sets of weak coherent states which span a tomographically complete set of MUBs to characterise the device. Afterwards, we use the retrieved measurement operators of our QPG to perform TM state tomography of randomly chosen TM states in an up to 7-dimensional Hilbert space with average fidelities of 0.99. This combines, for the first time, the necessary ingredients for high-dimensional QIS with single-photon TMs and paves the way towards future applications of this technology.

\section{Frequency conversion and mode selective measurements}
In this section, we present the theoretical basis behind the QPG and the use of it for tomography of TM states.
We express our single-photon states in terms of broadband TMs
\begin{equation} \hat{A}_i = \int f_i(\omega)\hat{a}(\omega) \mathrm{d} \omega, \label{eq:basis}\end{equation}
where $f_i(\omega)$ are frequency amplitudes and $\hat{a}(\omega)$ the annihilation operators for the central frequency $\omega$. 
The spectral intensity $|f_i(\omega)|^2$, can be measured with a standard spectrometer.
In the following, the modes $\hat{A}_i$ form our discrete basis of dimension $d$, i.e. the functions $f_i(\omega)$ are orthonormal and $0\leq i<d$. In the experiment, we take $d=\{ 5, 7\}$ and $f_i(\omega)$ as Hermite-Gaussian functions of order $i$.

\begin{figure}[t]
\begin{center}
\includegraphics[width=0.4\linewidth]{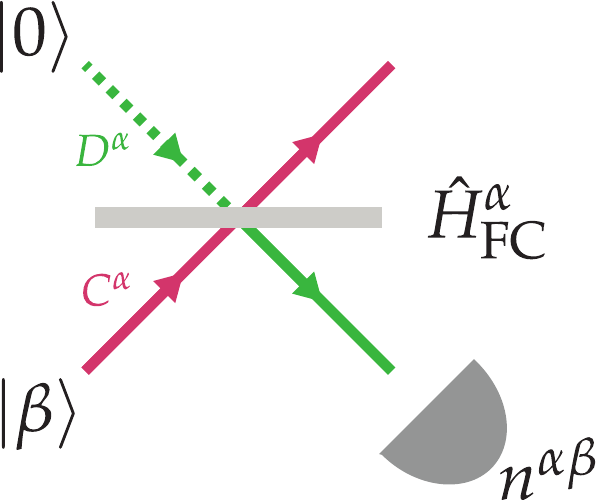}
\end{center}
\caption{Outline of QPG operation. The QPG is a beamsplitter operating on a TM defined by the index $\alpha$. For the measurement tomography, we send coherent states $\ket{\beta}$ to the QPG and at the converted (reflected) port we measure the number of converted photons using a bucket detector, noted as $n^{\alpha\beta}$.}
\label{fig:outline}
\end{figure}

Before giving the sketch of the TM tomography, we briefly review the underlying formalism of the QPG as a mode-selective frequency conversion (FC).
FC in general is a beam splitter acting on TMs, which is described by a Hamiltonian
$ \hat{H}_\mathrm{FC} = \theta \iint f^{\alpha}(\omega_\mathrm{in}, \omega_\mathrm{out}) \hat{a}(\omega_\mathrm{in}) \hat{b}^\dagger(\omega_\mathrm{out}) \mathrm{d}\omega_\mathrm{in} \mathrm{d}\omega_\mathrm{out} + \mathrm{h.c.}$, where $\hat{a}$ and $\hat{b}$ are the annihilation operators for the two beam-splitter modes. The transfer function 
\begin{equation} f^{\alpha}(\omega_\mathrm{in}, \omega_\mathrm{out}) = \alpha(\omega_\mathrm{pump})\Phi(\omega_\mathrm{in}, \omega_\mathrm{out}) 
\end{equation}
is given by the pump amplitude $\alpha(\omega_\mathrm{pump})$ and the phasematching function $\Phi(\omega_\mathrm{in}, \omega_\mathrm{out})$ of the crystal \cite{Brecht2011a, Eckstein2011}. 
We use an superscript $\alpha$ to indicate that we can adjust the process by shaping the pump spectrum. Using the Schmidt decomposition, the transfer function $f^\alpha(\omega_\mathrm{in}, \omega_\mathrm{out})$ can be decomposed into its eigenmodes defining new TM operators $\hat{C}^\alpha_k$ and $\hat{D}^\alpha_k$, thus reducing the integral to the following sum 
\begin{equation} \hat{H}^\alpha_\mathrm{FC} = \theta \sum_k \lambda_k^\alpha (\hat{D}^\alpha_k)^\dagger \hat{C}^\alpha_k + \mathrm{h.c.} , 
\end{equation}
where $\lambda_k^\alpha$ are the eigenvalues of the decomposition, normalised as $\sum_k |\lambda_k^\alpha|^2 = 1$, and $\theta$ is the gain of the process. The orthogonality of the eigenmodes ensures that we can regard the FC as independent beam splitters with a reflectivity or conversion efficiency of $\eta^\alpha_k = \sin^2(|\theta \lambda^\alpha_k|)$.
As sketched in Fig. \ref{fig:outline}, we have no input in mode $D$ and measure the mean photon number of the converted light, which is
\begin{equation} n = \sum_k \eta^\alpha_k \langle (\hat{C}^\alpha_k)^\dagger \hat{C}^\alpha_k \rangle. 
\end{equation}
To calculate what this means for a given input spectral shape, we decompose the mode $\beta$ of the input state into the eigenmodes of the FC
\begin{equation} \hat\beta = \sum_k v_k^{\alpha\beta} \hat{C}^\alpha_k. 
\end{equation}
Then we can rewrite the mean photon number of converted light as
\begin{equation} n^{\alpha\beta} = N^\beta \sum_k \eta_k^\alpha |v_k^{\alpha\beta}|^2, 
\end{equation}
where $N^\beta$  is the total mean photon number of the input state and $|v_k^{\alpha\beta}|^2$ the overlap between the input mode $\beta$ and the $k-$th eigenmode of the conversion process for a pump setting $\alpha$. Interestingly, this is valid for all photon number distributions including the coherent states we use here. 

We can also rewrite this in vector notation as
\begin{align}
n^{\alpha\beta} = N^\beta \sum_k \eta_k^\alpha |\langle \beta | k^\alpha \rangle |^2 
= \langle \beta | \hat{M}^\alpha | \beta  \rangle,
\end{align} 
where  $\hat{M}^\alpha = \sum_k \eta_k^\alpha | k^\alpha \rangle \langle k^\alpha | = \sum_{ij} m^\alpha_{ij} \ket{i} \bra{j}$ is our measurement operator, $\ket{i}$ the TM basis from Eq. (\ref{eq:basis}), $\ket{\beta}$ the input state and $\ket{k^\alpha}$ the eigenvectors of the process.
The idea of measurement tomography is to probe the matrix $\hat{M}^\alpha$ with different states $\ket{\beta}$. All we have to do is to generate a tomographically complete set of probe states and employ standard measurement tomography with the measured mean photon numbers for each setting, thus determining the elements $m_{ij}^\alpha$. Diagonalising this matrix, we get the FC eigenmodes $\ket{k^\alpha}$ and efficiencies $\eta_k^\alpha$. This fully characterises the input-mode structure of the FC. An ideal QPG has only one eigenmode, i.e. $\hat{M}^\alpha$ has only one non-zero eigenvalue, and the shape of the eigenmode would reflect the shape of the pump $k_0^\alpha(\omega) = \alpha(-\omega)$. This can be achieved in a three-wave mixing process with the group-velocity matching (GVM) condition between the input and the pump fields \cite{Brecht2011a,Eckstein2011}. 

It is worth noting that while the number of modes of the FC is in principle infinite, the probe space is only finite dimensional. Despite this, the reconstruction of the FC within the probe space is accurate. A simple example is when the TMs of the pump and input are not perfectly matched, e.g. in their central frequencies. 
This can change the overall conversion efficiency $\mathrm{tr} (\hat{M}^\alpha) = \sum_k \eta_k^\alpha$ for different pump shapes $\alpha$. We therefore try to match the central frequencies and bandwidths of the input and pump TMs to cover as much of the FC space as possible.

\section{Experiment}
\begin{figure}
\centering
\includegraphics[width=1\linewidth]{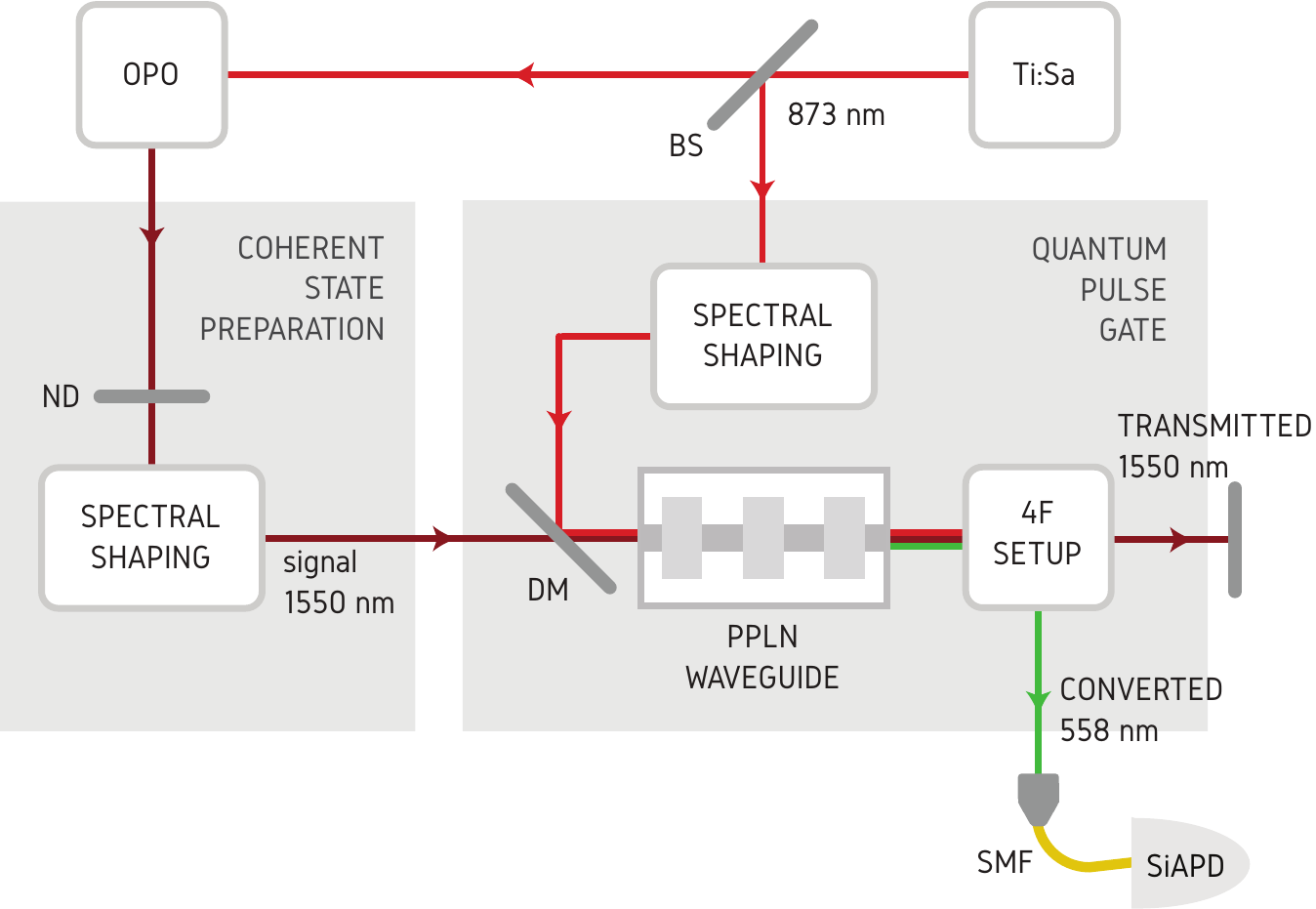}
\caption{Experimental setup. A femtosecond titanium:sapphire (Ti:Sa) oscillator with repetition rate of 80 MHz is used to pump an optical parametric oscillator (OPO). The pump of the QPG is obtained from a tap-off of Ti:Sa laser. The input signal field is prepared by attenuating the OPO output to a mean photon number of 0.1 photon per pulse by using neutral density (ND) filters. For spectral shaping, we use SLMs in a folded 4f-setup to shape the desired spectral amplitude and phase for the both fields. Then pump and input fields are combined on a dichroic mirror (DM) and coupled to a in-house built periodically poled lithium niobate (PPLN) waveguide, held at $207^{\circ}$C. After the PPLN waveguide, the up-converted photons with a green colour are selected by a 4f-setup and coupled to a silicon avalanche photodiode (SiAPD), through a single-mode fibre (SMF). }
\label{fig:experiment}
\end{figure}

The outline of the experimental setup is sketched in Fig. \ref{fig:experiment}. We take ultrashort pulses from a Ti:sapphire oscillator (Coherent Chameleon Ultra II) to pump an optical parametric oscillator (Coherent Chameleon OPO). With this configuration we have Gaussian pulses at central wavelengths of 873 nm and 1550 nm, for the pump and signal fields respectively, with amplitude FWHM of 3.35 THz for both fields. To prepare the coherent input state, we attenuate the OPO beam to a mean photon number of 0.1 per pulse. We use a self-built pulse shaper to shape the pump and a commercial pulse shaper (Finisar waveshaper 4000) to shape the input light pulses, with spectral resolutions of 22 pm and 8 pm respectively. 
The self-built pulse shaper is a folded 4f-setup consist of a magnifying telescope, a holographic diffraction grating with 2000 lines per mm, a cylindrical silver mirror and a reflective liquid crystal on silicon spatial light modulator (Hamamatsu X10468-07 LCoS-SLM).
We use spectral interferometry to ensure both pulse shapers are dispersion free. The shaping resolutions are better than the resolution we require in this experiment. For example, while we could prepare the 20\textsuperscript{th}-order Hermite-Gaussian mode, we only use the first 7 modes as our basis due to other constraints that will be discussed later. For the tomography, we choose a bandwidth of 0.4 THz (FWHM of the amplitude of the Gaussian mode) for both fields. Finally, the type-II sum-frequency process happens in an in-house built 17 mm LiNbO\textsubscript{3} crystal with titanium indiffused waveguides and a poling period of 4.4 $\mu$m. The waveguides are designed to be spatially single mode at 1550 nm. 

\begin{figure}
\centering
\includegraphics[width=1\linewidth]{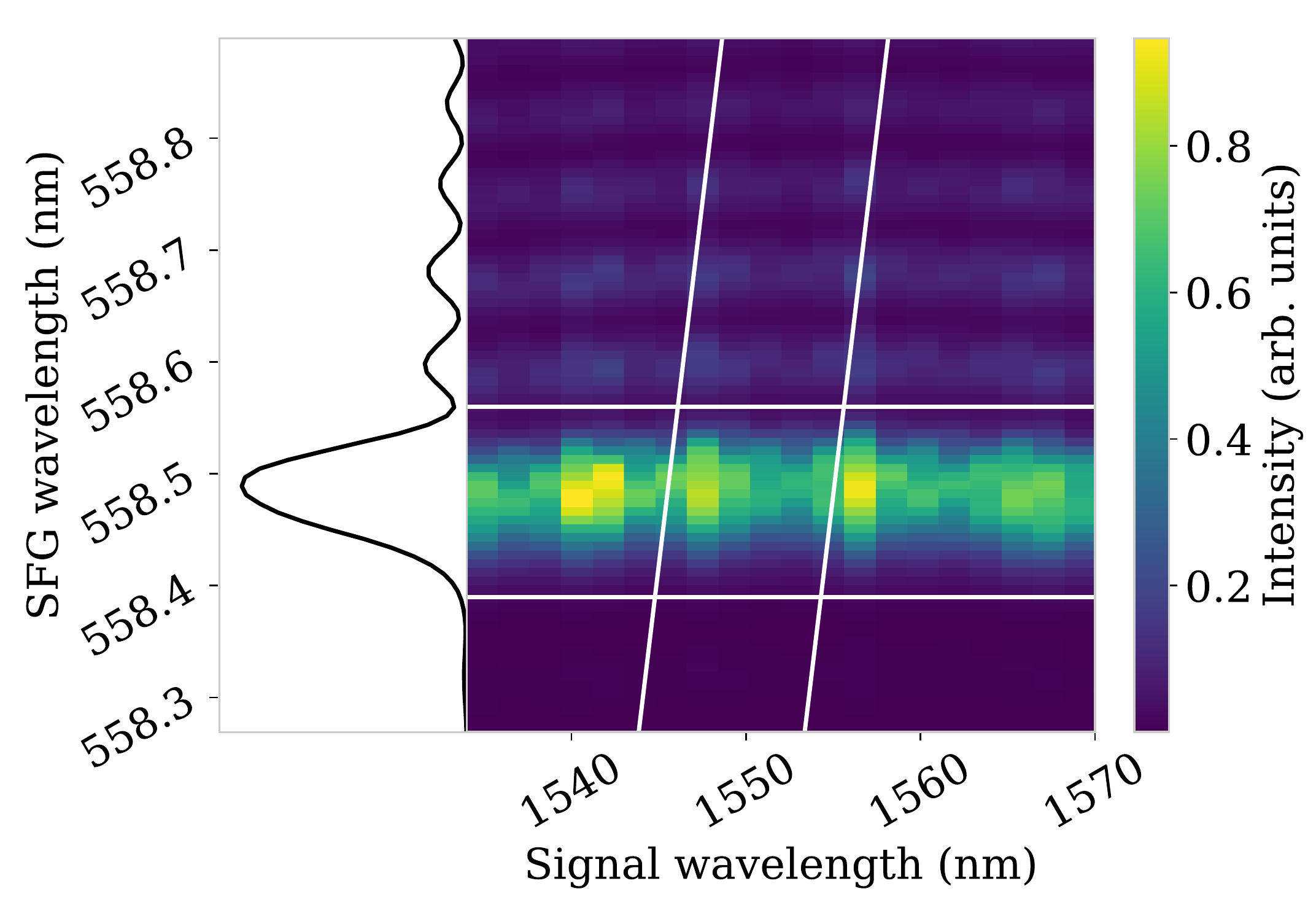}
\caption{Phasematching function of the QPG. Right: The zero gradient of the phasematching function $\Phi(\omega_\mathrm{in}, \omega_\mathrm{out})$, is an indicator of group-velocity matching between input signal and the pump field. The diagonal white lines are marking the orientation of the pump amplitude $\alpha(\omega_\mathrm{out}-\omega_\mathrm{in})$ and the bandwidth we use in this paper. The horizontal white lines are showing the bandwidth of the 4f-setup used to filter the SFG signal. Left: marginal distribution of the plot on the right side. Asymmetries are due to inhomogeneity of the effective refractive index along the waveguide.}
\label{fig:pm}
\end{figure}

The key property of a QPG is the group-velocity matching (GVM) between the input and the pump \cite{Brecht2011a,Eckstein2011}. In Fig. \ref{fig:pm} we plot the intensity of the phasematching function $|\Phi(\lambda_\mathrm{in}, \lambda_\mathrm{out})|^2$, measured with a scanning continuous wave input laser and adjusted pump pulses on a high-resolution spectrometer. A perfect GVM condition results in zero gradient of the phasematching function in Fig. \ref{fig:pm}.
The marginal spectrum of this function, plotted on the left side in Fig. \ref{fig:pm}, shows an asymmetric structure with decaying side lobes. This can be explained by an inhomogeneity of the effective refractive index along the waveguide, equivalent to a variation of the poling period. A quadratic variation of the poling period can introduce such asymmetric side peaks. In the experiment, we also have a 4f-setup on the SFG line (with a total transmissivity of about 0.55) that allows us to filter out these side lobes.

One common complication with waveguides is that different spatial mode combinations have different phasematchings. In our case, these do not overlap with the phasematching for the fundamental mode shown in Fig. \ref{fig:pm}, thus we can simply filter them out spectrally. Nevertheless, special care is taken to optimise the coupling of both beams into the waveguide for the desired process and minimise the intensity of higher order modes.

We shape both the pump and the input to span a complete set of mutually unbiased bases (MUBs) \cite{Bandyopadhyay2002}. These have the property that for a dimension $d$, there are $(d+1)$ bases such that overlaps between states from different bases are always $1/d$, hence unbiased. This ensures that the space is uniformly probed. Furthermore, the total set is tomographically over-complete, helping to reduce systematic experimental errors. Since for each pump shape, we have to run the full characterisation with $(d+1)d$ input modes, the total number of measurements for $d=5$ and $d=7$ are 900 and 3136, respectively. For each of them, we record counts for about 1 s at count rates up to $10^5$ counts/s. This corresponds to a FC efficiency of about five percent, which is solely limited by the pump pulse energy of about 5 pJ in the our current experimental setup. Despite the relatively low conversion efficiency, a short measurement time is possible owing to high detection efficiency of the SiAPD. Since the count rates are directly proportional to the powers of the pump and the input, we record both values after the waveguide and normalise the count rates accordingly to account for small drifts in the setup (with the magnitude of less than 10\%). It worth mentioning that one can also use symmetric informationally complete POVMs (SIC-POVMs) as the tomography bases \cite{Scott2010}. The main advantage of the SIC-POVMs is that, contrary to MUBs, they exist for any arbitrary dimension \cite{Renes2004}.

\section{Measurement tomography of the QPG}
\begin{figure*}
\centerline{\includegraphics[width=1\linewidth]{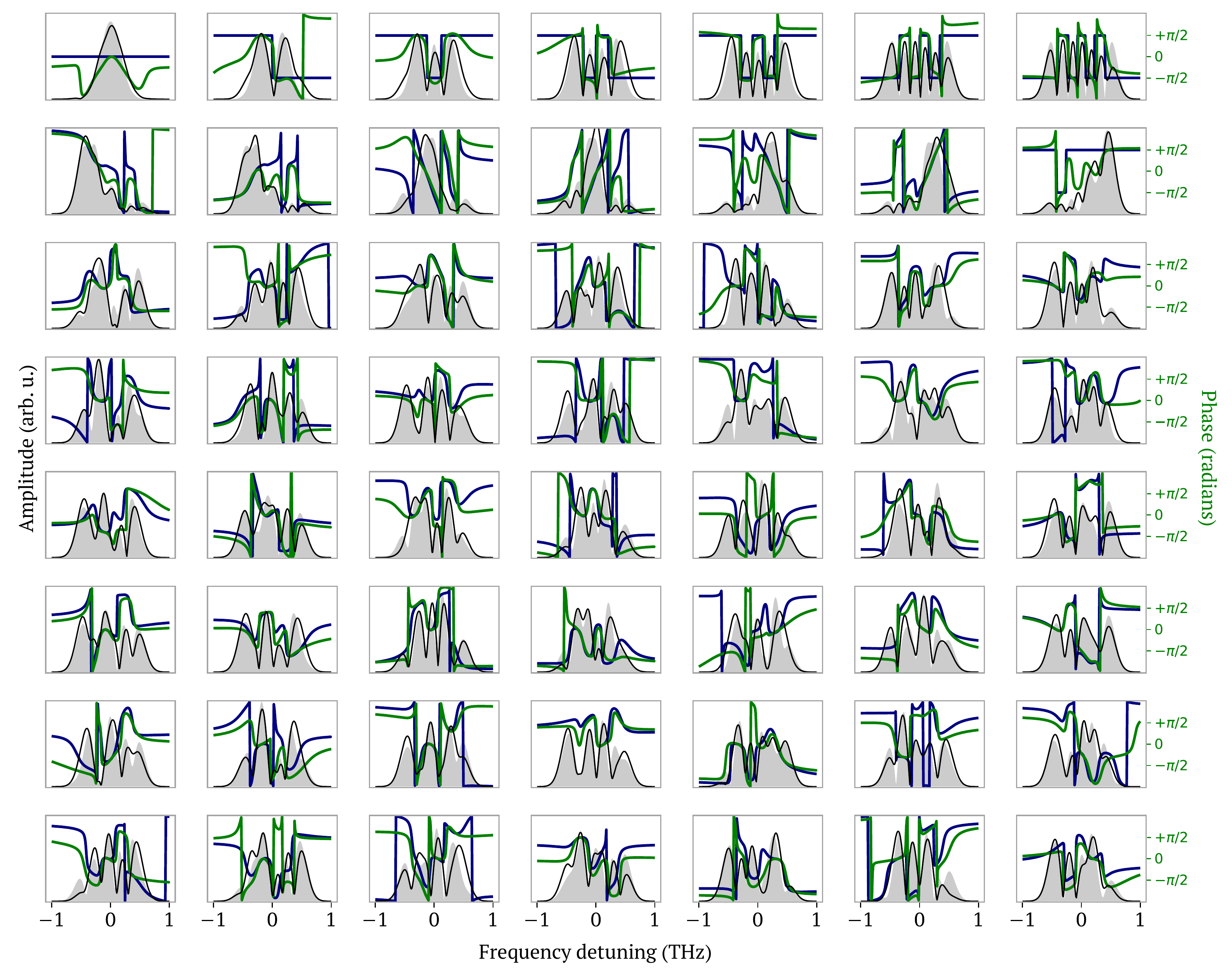}}
\caption{The first eigenvectors of the 7$\times$6 measurement operators. For each plot, the x-axis corresponds to the frequency detuning (from the central frequency) and the y-axis to the amplitude and phase. Black and green lines are the measured amplitudes and phases, respectively; shaded areas and blue lines correspond to the theoretical MUB modes. Note that the phase is $2\pi$ periodic, which is also the interval of the y-axis. Please note that phases are only meaningful when a significant amplitude is present. 
}
\label{fig:M}
\end{figure*}

To find the measurement operators $\hat{M}^\alpha$ from the data we perform a weighted least squares fit
\begin{equation} \underset{\hat{M}^\alpha}{\mathrm{min}} \sum_\beta \frac{|f^{\alpha \beta} - \langle \beta | \hat{M}^\alpha | \beta \rangle |^2}{f^{\alpha \beta}}, \end{equation}
where $f^{\alpha \beta}$ are normalised count rates and $\hat{M}^\alpha$ is constrained to be Hermitian and positive semidefinite. Since each setting $\alpha$ is an independent measurement, we do not put a constraint on the sum of operators. 
In Fig. \ref{fig:M} we show the first eigenmodes of all measurement operators for 7 dimensions. They closely resemble the ideal MUB states. 
Additionally, the matrix of projections of MUB POVM elements which shows the orthogonality of the basis is given in Appendix \ref{sec:mub}.

To quantify how accurate the results are, we calculate the purities $\mathcal{P}^\alpha = \mathrm{tr}([\hat{M}^\alpha]^2) / \mathrm{tr}(\hat{M}^\alpha)^2 $ and the fidelities $\mathcal{F}^\alpha =\sqrt{\bra{\alpha} \hat{M}^\alpha \ket{\alpha} / \mathrm{tr}(\hat{M}^\alpha)}$ with the ideal operators $| \alpha \rangle \langle \alpha |$. We perform the characterisation in 5 and 7 dimensions, whereas for 5 dimensions we also compare the two experimental settings with and without a spectral filter in the output mode. As mentioned, the spectral filter blocks the side lobes of the phasematching. 
The average values with their respective standard deviations are listed in Table \ref{tab:detTom}.
For comparison we also show theoretical values assuming a Gaussian horizontal phasematching and perfect pump shaping. The imperfections in this case originate from the fact that the phasematching is only about five times narrower than the pump, leading to correlations in the transfer function and multimode performance of the QPG. These correlations also explains why suppressing the side lobes of the output spectrum improves the purity from 0.72 to 0.92. A comparison of the eigenmodes for these two cases shows that the first eigenmode hardly changes. Thus the spectral filtering suppresses the higher order spectral modes introduced by the side lobes of the phasematching, or in other words drives the QPG closer to single-modeness. Going from 5 to 7 dimensions slightly lowers both the purities and the fidelities. One reason is that the richer spectral structure of the pump at higher dimensions, again, will introduce some spectral correlations in the transfer function which also reduce the theoretical values. However, the expected reduction is smaller than what we measure. Imperfections in the pulse shaping are a greater problem for higher dimensions. With the increase of dimensionality, the total bandwidth both in time and frequency increases which requires the relative phases and amplitudes to be accurate over a broader range in both time and frequency. To improve the single-mode operation of QPG, one can use a longer waveguide which gives a narrower phasematching bandwidth. Furthermore, the measurement time increases drastically which makes the experiment more susceptible to drifts in the setup. With the current experimental setup, the 7 dimensional characterisation takes about 2 hours.

\begin{table}
\centering
\caption{Purities and fidelities of QPG measurement operators.}
\begin{tabular}{lccc}
\hline
d & 5 (unfiltered) & 5 & 7 \\
\hline
$\mathcal{P}_\mathrm{measured}$ &$0.719 \pm 0.064$ \ \ & $0.920 \pm 0.024$ \ \ & $0.811 \pm 0.035$ \\
$\mathcal{F}_\mathrm{measured}$ &$0.778 \pm 0.086$ \ \ & $0.912 \pm 0.046$ \ \ & $0.847 \pm 0.042$ \\
$\mathcal{P}_\mathrm{theory}$ &  & $0.939 \pm 0.026$ \ \ & $0.909 \pm 0.035$ \\
$\mathcal{F}_\mathrm{theory}$ &  & $0.979 \pm 0.008$ \ \ & $0.971 \pm 0.010$ \\
\hline
\end{tabular}
  \label{tab:detTom}
\end{table}

The overall high fidelities we measure in this work demonstrates that the QPG can operate on arbitrary TMs in a selective way. 
The fidelities also quantify the mode selectivity since the normalised conversion efficiency is given by $\mathcal{F}^2$. In the 5-dimensional case, that means that the desired mode gets converted with 83\% efficiency and any orthogonal mode gets converted with less than 17\%. However, with the measurement operators we have much more information than just the mode selectivity. For a task like state tomography, the QPG operation can be calibrated for small experimental errors, as we have here. All we need is mode sensitivity and the knowledge of our mode detector, which we have with the matrices $\hat{M}^\alpha$. In Appendix \ref{sec:scalability} we discuss the feasibility of this tomographic method at the presence of more significant experimental errors.


\begin{figure*}[htbp]
\centering
\includegraphics[width=.9\linewidth]{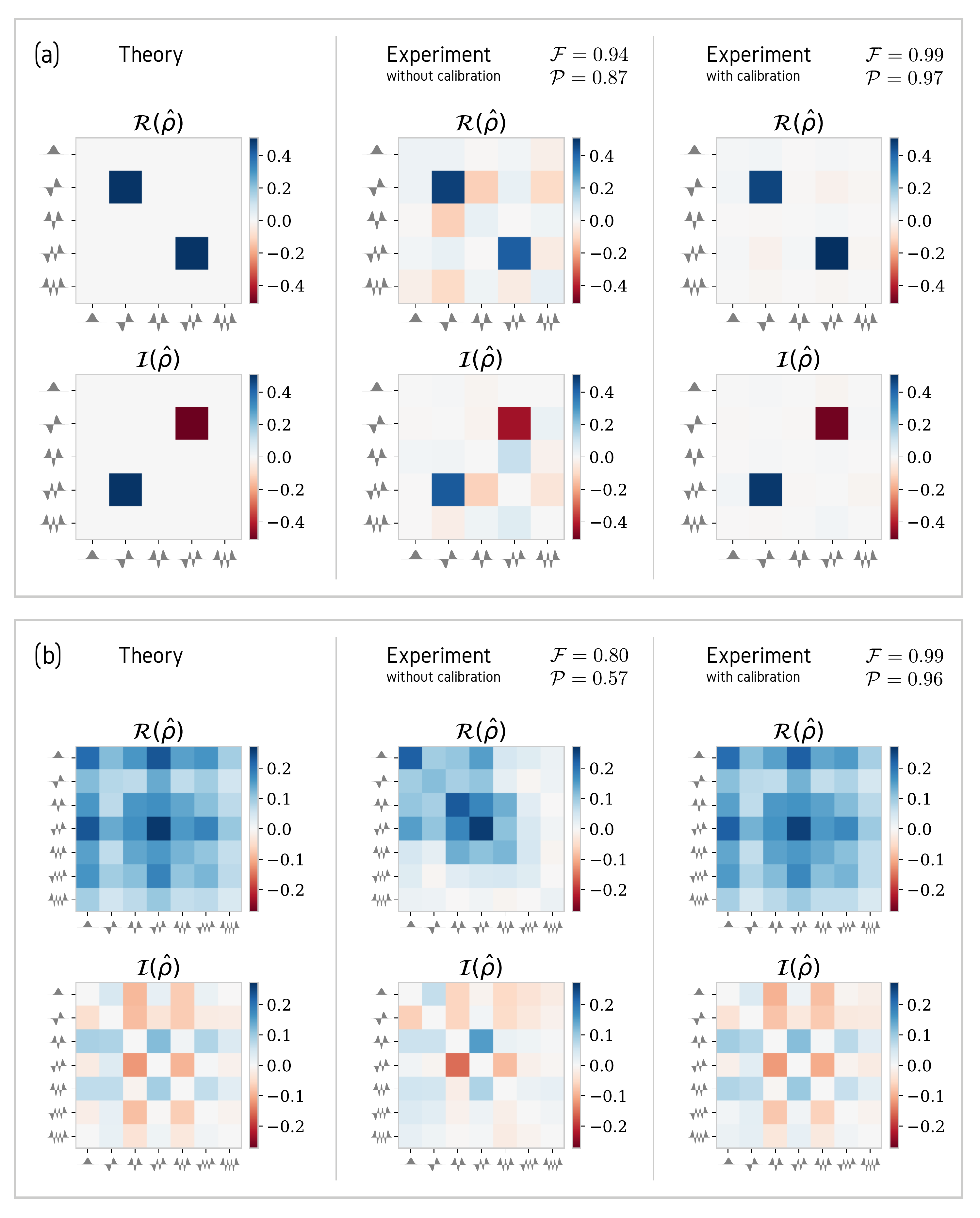}
\caption{Two examples of state tomography with QPG in the Hermite-Gaussian basis in five (a) and seven (b) dimensions. State vectors corresponding to each density matrix is detailed in the Appendix \ref{sec:states}. For each state the theoretical density matrix (left), the reconstructed density matrix without QPG calibration (middle) and the reconstructed density matrix with QPG calibration (right) are plotted.}
\label{fig:state}
\end{figure*}

\section{State tomography with the QPG}
In the following we investigate the performance of the QPG for state tomography. For this purpose, we prepare states like $\hat\rho = \ket{\beta}\bra{\beta}$, which are different from the characterisation set we use for the measurement tomography. To ensure fair benchmarking we prepare twenty different input states where half of them are generated randomly. Then we use the $(d+1)d$ QPG settings $\alpha$ to reconstruct the input state. We measure the normalised probabilities $f^\alpha$ and minimise 
\begin{equation} 
\underset{\hat\rho}{\mathrm{min}} \sum_\alpha \frac{|f^{\alpha} - \mathrm{tr}(\hat\rho \hat{M}^\alpha) |^2}{f^{\alpha}} \quad, 
\label{eq:minState}
\end{equation}
under the constraints that $\hat\rho$ is Hermitian, positive semidefinite and $\mathrm{tr}(\hat\rho)=1$.
First, we assume a perfect QPG with ideal measurement operators and reconstruct the input states. Since the prepared inputs are coherent states in well-defined TMs, we expect to reconstruct pure states. The average fidelities and their standard deviations measured for all input states are listed in Table \ref{tab:stateTom}, which shows a modest fidelity of the reconstructed state with respect to the prepared state. This is because the slight multimodeness of the QPG operation, translates into the mixedness of the reconstructed states and leads to inaccurate tomography.

\begin{table}[t]
\centering
\caption{Measured purities and fidelities of state tomography.}
\begin{tabular}{cccc}
\hline
d & 5 (unfiltered) & 5 & 7 \\
\hline
$\mathcal{P}$ \ \ & $0.68 \pm 0.079$ \ \ & $0.753\pm 0.098$ \ \ & $0.619 \pm 0.052$ \\
$\mathcal{F}$ \ \ & $0.742 \pm 0.126$ \ \ & $0.879\pm 0.041$ \ \ & $0.813 \pm 0.031$ \\
\hline
\end{tabular}
  \label{tab:stateTom}
\end{table}

\begin{table}[t]
\centering
\caption{Measured purities and fidelities of state tomography with calibrated QPG.}
\begin{tabular}{cccc}
\hline
d & 5 (unfiltered) & 5 & 7 \\
\hline
$\mathcal{P}$ & $0.931 \pm 0.038$ \ \ & $0.972 \pm 0.016$ \ \ & $0.957 \pm 0.017$ \\
$\mathcal{F}$ & $0.971 \pm 0.015$ \ \ & $0.991 \pm 0.005$ \ \ & $0.988 \pm 0.004$ \\
\hline
\end{tabular}
  \label{tab:stateTomCalib}
\end{table}

To improve the quality of the state tomography we can use the characterised measurement operators of the QPG in Eq. (\ref{eq:minState}). Table \ref{tab:stateTomCalib} summarises the outcome. The improvement is striking. We obtain fidelities of 0.99 with the actual input state. Two example of such states are shown in Fig. \ref{fig:state}. The decrease in fidelity from 5 to 7 dimensions is almost negligible and even without filtering, the values are still very high. This shows the power of proper detector calibration for state tomography. The outstanding fidelities suggest that the state tomography with QPG can be scaled up to higher dimensions. However performing a complete measurement tomography for higher dimensions, with the current experimental configuration, would require an impractically long measurement time. This is primarily a technical challenge to decrease the switching time of the SLMs and increase the count rates per second. From the numeric point of view, measurement tomography becomes time consuming very quickly. Here, one could switch to pattern tomography \cite{Rehacek2010}, which circumvents this tedious step by fitting the detector response pattern directly. We tested this approach as well and obtained similar fidelities as shown in Table \ref{tab:stateTomCalib}.

\section{Conclusion}
In conclusion, we experimentally characterised the measurement operators of a temporal-mode selective device in up to seven dimensions. We have shown that the device is effective in superposition bases spanning a tomographically complete set of mutually unbiased bases. Furthermore, we have shown that characterisation of the measurement operators of such a device enables accurate temporal-mode state tomography, with fidelities in the 0.99 range. With such characterisation, the QPG can be used to fully characterise ultrafast quantum states. Future work will focus on improving the performance of the QPG to realise its full potential for high-dimensional quantum information science with temporal modes.

\begin{acknowledgments}
The authors would like to thank Saleh Rahimi-Keshari, John Donohue, Nicolas Treps, Jonathan Roslund, and Dileep V. Reddy for helpful discussions. This research has received funding from the Gottfried Wilhelm Leibniz-Preis and from the European Union’s Horizon 2020 research and innovation programme under grant agreement No 665148.
\end{acknowledgments}

\appendix
\section{List of state vectors}
\label{sec:states}
The following is the list of state vectors associated with the density matrices presented in Fig \ref{fig:state}, described in the Hermite-Gaussian basis.

\begin{equation}
\ket{\psi_a} = \ket{1}-\imath\ket{3}
\end{equation}

\begin{equation}
\begin{aligned}
\ket{\psi_b} = & (0.36110833+0.28107443\imath)\ket{0} \\& + (0.14599764+0.23536858\imath)\ket{1} \\& + (0.39339517+0.05872998\imath)\ket{2} \\& +
  (0.37242591+0.35380667\imath)\ket{3} \\& + (0.34693250+0.07796563\imath)\ket{4} \\& + (0.25172264+0.24799887\imath)\ket{5} \\& +
  (0.16147789+0.12004762\imath)\ket{6}
\end{aligned}
\end{equation}

\section{MUB POVMs orthogonality}
\label{sec:mub}
\begin{figure}[htbp]
\centering
\includegraphics[width=.85\linewidth]{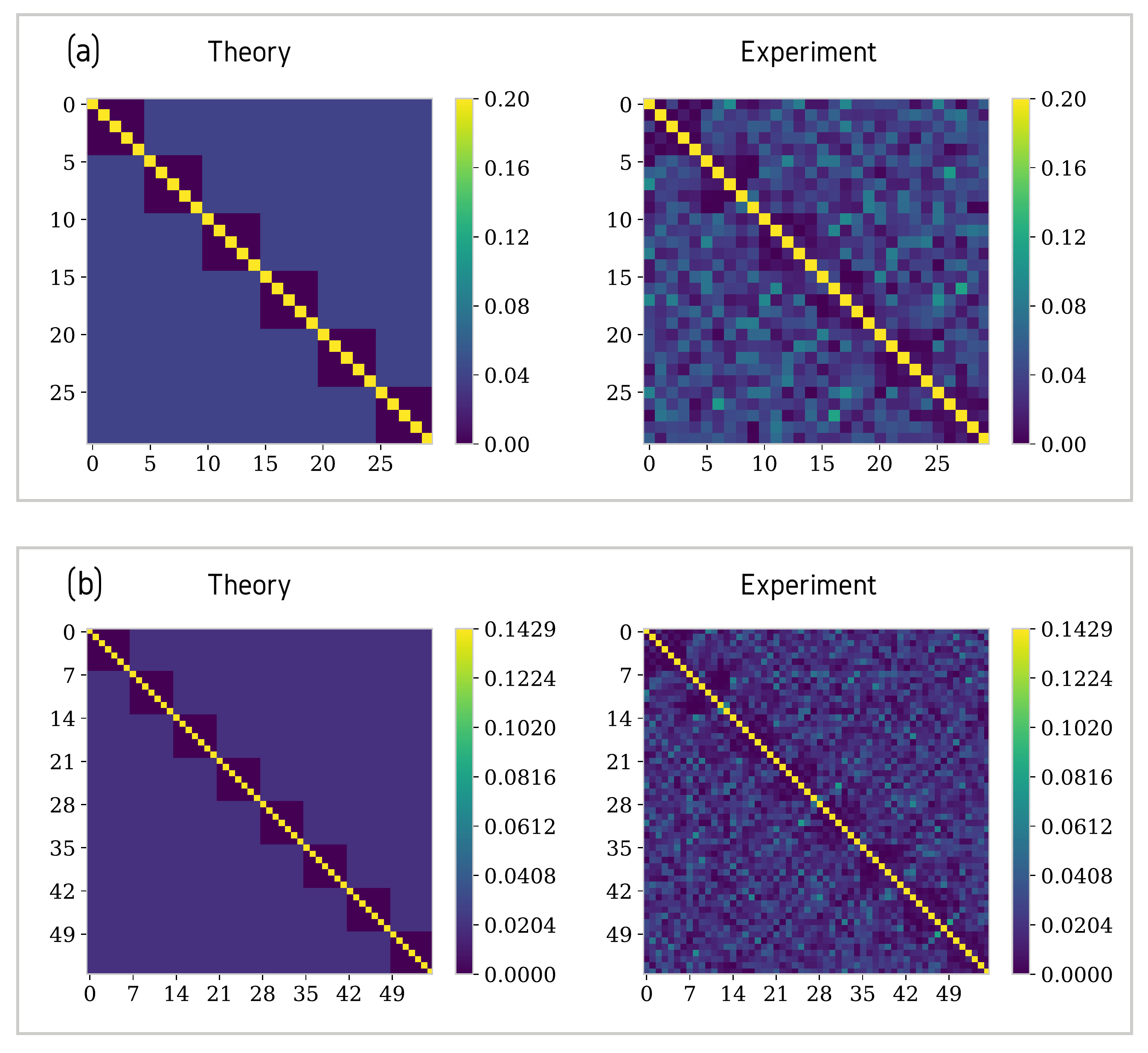}
\caption{Matrix of projections of MUB POVM elements for five (a) and seven (b) dimensions.}
\label{fig:orth}
\end{figure}

Fig. \ref{fig:orth} shows the matrix of projections of MUB POVM elements $|\ket{\phi_{i}}\bra{\phi_{j}}|^2$ for five and seven dimensions, which is used for normalising the data.

\section{Feasibility against experimental errors}
\label{sec:scalability}
\begin{figure}[htbp]
\centering
\includegraphics[width=1\linewidth]{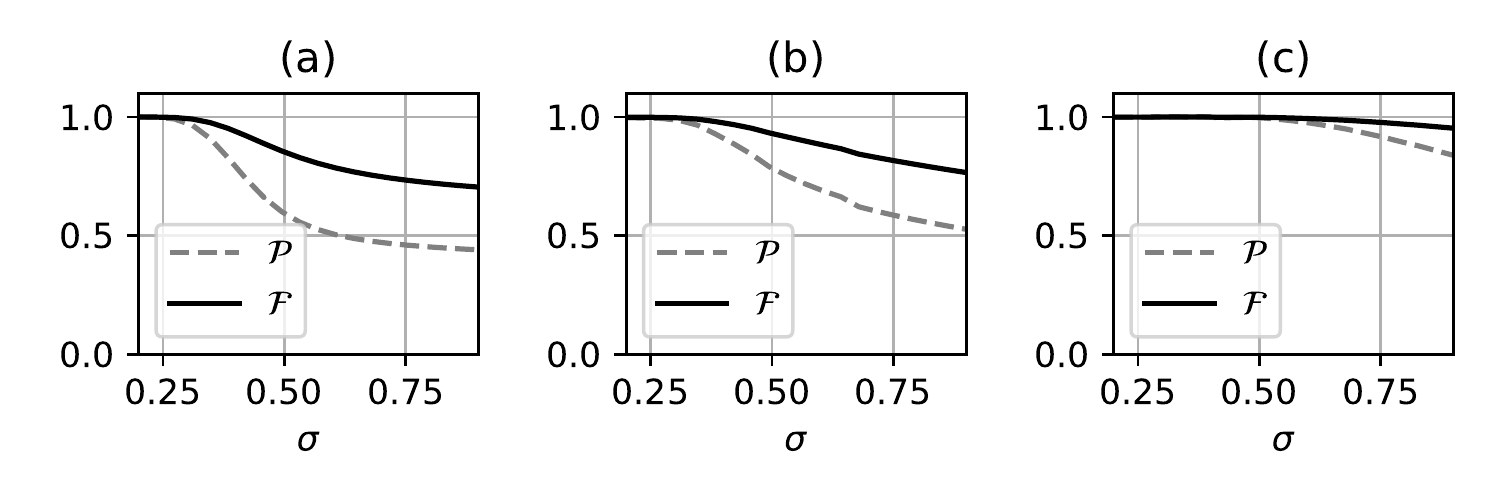}
\caption{The impact of an imperfect QPG (parametrised in $\sigma$ with arbitrary units) on fidelity and purity of (a) measurement operators, (b) state tomography without calibration, and (c) state tomography with calibration of the POVMs.}
\label{fig:scal1}
\end{figure}

In this section we briefly discuss the effectiveness of the our tomographic method against the imperfections of the QPG's measurement operators. To simulate the imperfect QPG measurements, we convolve the theoretical seven-dimensional measurement operators with a Gaussian filter with a width of $\sigma$. This error model is chosen because from operational point of view the main source of errors is the imperfect mode selectivity of QPG. With an increasing width of the Gaussian filter, purity and fidelity of the measurement operators decline, as plotted in Fig. \ref{fig:scal1} (a). In Fig. \ref{fig:scal1} (b), we use these imperfect measurement operators to perform a state tomography on a pure input state in the Gaussian mode; in which as expected, shows a reduced fidelity with increasing values of $\sigma$. Finally, in Fig. \ref{fig:scal1} (c), we use our knowledge of imperfect POVMs and repeat the state tomography with a calibrated QPG. For relatively small values of $\sigma$, with the purity of the measurement operators larger than about 0.6, the state tomography works with very high fidelities. However, our method breaks down for a larger amount of errors; which is considerably more than the experimental imperfections presented in this manuscript. With an excessive amount of experimental errors, other tomographic methods, such as Bayesian mean estimation \cite{Schmied2016}, might be more effective. Nonetheless, a comprehensive theoretical evaluation of various types of error and finding the optimised tomographic method is necessary, which is beyond the scope of this paper.

%

\end{document}